\documentclass[twocolumn,amsmath,nofootinbib]{revtex4} 

\usepackage{url}
\usepackage{amssymb}
\usepackage{amsfonts}
\usepackage{amsbsy}
\usepackage{graphicx}
\usepackage{hyperref}
\usepackage{array} 
\usepackage{multirow}

\newcolumntype{x}[1]{%
>{\centering\hspace{0pt}}p{#1}}%
\providecommand{\openone}{\leavevmode\hbox{\small1\kern-3.8pt\normalsize1}}

\def\ie{{\frenchspacing\it i.e.}}
\def\eg{{\frenchspacing\it e.g.}}

\def\spose#1{\hbox to 0pt{#1\hss}}
\def\simlt{\mathrel{\spose{\lower 3pt\hbox{$\mathchar"218$}}
   \raise 2.0pt\hbox{$\mathchar"13C$}}}
\def\simgt{\mathrel{\spose{\lower 3pt\hbox{$\mathchar"218$}}
     \raise 2.0pt\hbox{$\mathchar"13E$}}}
 \def\simpropto{\mathrel{\spose{\lower 3pt\hbox{$\mathchar"218$}}
     \raise 2.0pt\hbox{$\propto$}}}

\def\beq#1{\begin{equation}\label{#1}}
\def\eeq{\end{equation}}
\def\beqa#1{\begin{eqnarray}\label{#1}}
\def\eeqa{\end{eqnarray}}
\def\eq#1{equation~(\ref{#1})}

\def\fig#1{Figure~\ref{#1}}
\def\Fig#1{Figure~\ref{#1}}

\def\Sec#1{Section~\ref{#1}}

\def\ed{\end{document}}

\def\p{{\bf p}}
\def\n{{\bf n}}
\def\q{{\bf q}}
\def\r{{\bf r}}
\def\s{{\bf s}}

\def\u{{\bf u}}
\def\vv{{\bf v}}
\def\w{{\bf w}}
\def\x{{\bf x}}
\def\y{{\bf y}}
\def\z{{\bf z}}

\def\bzero{{\bf 0}}

\def\I{{\bf I}}
\def\A{{\bf A}}
\def\B{{\bf B}}
\def\C{{\bf C}}
\def\D{{\bf D}}

\def\F{{\bf F}}
\def\G{{\bf G}}
\def\GG{{\bf\Gamma}}

\def\K{{\bf K}}
\def\Lamb{{\bf\Lambda}}

\def\P{{\bf P}}
\def\PP{{\mathbf\Pi}}
\def\Q{{\bf Q}}
\def\R{{\bf R}}
\def\Re{\mathbb{R}}

\def\Sig{{\bf\Sigma}}
\def\T{{\bf T}}
\def\U{{\bf U}}

\def\V{{\bf V}}

\def\diag{\hbox{diag}\,}

\def\expec#1{\langle#1\rangle}

\def\rn{}
\def\nn#1 #2{#2. #1}				
\def\nnn#1 #2 #3{#2. #3. #1}			
\def\nnnn#1 #2 #3 #4{#2. #3. #4 #1}		
\def\nnnnn#1 #2 #3 #4 #5{#2. #3. #4 #5. #1}	

\def\rf#1;#2;#3;#4;#5 {{\frenchspacing\par\rn#1, #3 {\bf #4}, #5 (#2). \par}}
\def\rg#1;#2;#3;#4;#5;#6 {{\frenchspacing\par\rn#1, #3 {\bf #4}, #5 (#2). \par}}
\def\rfbook#1;#2;#3;#4;#5 {{\frenchspacing\par\rn#1, {\it #3} (#5, #4, #2).\par}}
\def\rfprep#1;#2;#3 {{\par\frenchspacing\rn#1, #3 (#2).\par}}
\def\rfproc#1;#2;#3;#4;#5;#6 {{\frenchspacing\par\rn#1 #2, in {\it #3}, ed. #4 (#5: #6)\par}}
\def\rfprocp#1;#2;#3;#4;#5;#6;#7 {{\frenchspacing\par\rn#1 #2, in {\it #3}, ed. #4 (#5: #6), p#7\par}}

\begin{document}
\pdfoptionalwaysusepdfpagebox=5


\title{Latent Representations of Dynamical Systems: When Two is Better Than One}

\author{Max Tegmark}

\address{Dept.~of Physics, MIT Kavli Institute \& Center for Brains, Minds \& Machines, Massachusetts Institute of Technology, Cambridge, MA 02139; tegmark@mit.edu}

\date{\today}


\begin{abstract}
A popular approach for predicting the future of dynamical systems involves mapping them into a lower-dimensional ``latent space'' where prediction is easier. We show that the information-theoretically optimal approach uses different mappings for present and future, in contrast to state-of-the-art machine-learning approaches where both mappings are the same. We illustrate this dichotomy by predicting the time-evolution of coupled harmonic oscillators with dissipation and thermal noise, showing how the optimal 2-mapping method significantly outperforms principal component analysis and all other approaches that use a single latent representation, and discuss the intuitive reason why two representations are better than one. We conjecture that a single latent representation is optimal only for time-reversible processes, not for {\eg} text, speech, music or out-of-equilibrium physical systems.
\end{abstract}

\maketitle


\section{Introduction}

A core challenge in physics (and in life quite generally) is data distillation: 
keeping only a manageably small fraction of our available data that nonetheless retains 
most of the information that is useful to us. Ideally, the information can be partitioned into a set of independent chunks and sorted from most to least useful, enabling us to select the number of chunks to retain so as to optimize our tradeoff between utility and and data size.

Consider a random vector $\x$, and partition its elements into two parts:
\beq{xDefEq}
\x=\left({\x_1\atop\x_2}\right).
\eeq
We may, for example, interpret the vectors $\x_1$ and $\x_2$ as observations of two separate systems at the same time, 
or as two separate observations of the same system some fixed time interval $\Delta t$ apart. 
Let us now consider various forms of ideal data distillation, as summarized in Table~\ref{ComparisonTable}.

\begin{table}[h!]
\begin{tabular}{|c|l|l|l|}
\hline
Random		&What is				&\multicolumn{2}{c|}{Probability distribution}\\
\cline{3-4}
vectors		&distilled?			&Gaussian			&Non-Gaussian\\
\hline
1			&Entropy				&PCA				&Autoencoder\\
			&$H(\x)=\sum H(u_i)$	&$\u=\F\x$			&\u=f(\x)\\
\hline
2			&Mutual information		&CCA				&Latent reps\\
			&$I(\x,\y)=\sum I(u_i,v_i)$	&$\u=\F\x$			&$\u=f(\x)$\\
			&					&$\vv=\G\y$			&$\vv=g(\y)$\\
\hline
\end{tabular}
\caption{Data distillation: the relationship between Principal Component Analysis (PCA), Canonical Correlation Analysis (CCA), nonlinear autoencoders and nonlinear latent representations.
\label{ComparisonTable}
}
\end{table}

If we distill $\x$ as a whole, then we would ideally like to find a function $f$ such that
the so-called latent representation $\u=f(\x)$ retains the full entropy
$H(\x)=H(\u)=\sum H(u_i)$, decomposed into independent\footnote{When implementing any distillation algorithm in practice, there is always a one-parameter tradeoff between compression and information retention which defines a Pareto frontier. A key advantage of the latent variables (or variable pairs) being statistically independent is that this allows the Pareto frontier to be trivially computed, by simply sorting them by decreasing information content and varying the number retained.
}
parts with vanishing mutual infomation: 
$I(u_i,u_j)=\delta_{ij}H(u_i).$
For the special case where $\x$ has a multivariate Gaussian distribution, 
the optimal solution is 
Principal Component Analysis (PCA) \cite{pearsonPCA1901}, which
has long been a workhorse of statistical physics and many other disciplines: here $f$ is simply a linear function mapping into the eigenbasis of the covariance matrix of $\x$.
The general case remains unsolved, and it is easy to see that it is hard: 
if $\x=c(\u)$ where $c$ implements some state-of-the-art cryptographic code, then finding $f=c^{-1}$ (to recover the independent pieces of information and discard the useless parts) would generically require breaking the code. Great progress has nonetheless been made for many special cases, using techniques such as nonlinear autoencoders \cite{vincent2008extracting} and Generative Adversarial Networks (GANs) \cite{goodfellow2014generative}.

Now consider the case where we wish to distill $\x_1$ and $\x_2$ separately,
into $\u\equiv f(\x_1)$ and $\vv=g(\x_2)$, retaining the mutual information between the two parts. 
Then we ideally have
$I(\x,\y)=\sum I(u_i,v_i)$,
$I(u_i,u_j)=\delta_{ij}H(u_i),$
$I(v_i,v_j)=\delta_{ij}H(v_i),$
$I(u_i,v_j)=\delta_{ij}I(u_i,v_i).$
This problem has attracted great interest, especially for time series where $\x_1=\z_i$ and $\x_2=\z_j$ for some sequence of states $\z_k$ ($k=0,1, 2, ...$) in physics or other fields, 
where one typically maps the state vectors $\z_i$ into some lower-dimensional 
vectors $f(\z_i)$, after which the prediction is carried out in this latent space. 
For the special case where $\x$ has a multivariate Gaussian distribution, 
the optimal solution is 
Canonical Correlation Analysis (CCA) \cite{hotellingCCA1936}: here both $f$ and $g$ are linear functions, computed via a  
singular-value decomposition (SVD) \cite{eckart1936SVD} of the cross-correlation matrix after prewhitening $\x_1$ and $\x_2$.
The general case remains unsolved, and is obviously even harder than the above-mentioned 1-vector autoencoding problem. 
 The recent DeepMind paper \cite{oord2018representation} reviews the state-of-the art as well as presenting  Contrastive Predictive Coding, a powerful new distillation technique for time series, following the long tradition of setting $f=g$.

The purpose of this paper is to further investigate the case for choosing $f\ne g$. 
We will do this by studying the lower-left quadrant of Table~\ref{ComparisonTable}, where information-theoretically optional results can be derived, and using these results to discuss implications for the harder problem in the lower-right quadrant.
The rest of this paper is organized as follows. \Sec{LatentSec} discusses analytic results for the lower-left quadrant. In \Sec{OscilatorSec}, the optimal $f\ne g$ method is benchmarked on a physics example, showing significant improvement over $f=g$ methods. Our conclusions are discussed in \Sec{ConclusionsSec}.

\section{CCA implications for latent representations: two is better than one}
\label{LatentSec}

\subsection{Notation}

Without loss of generality, we take the random vector $\x$ from \eq{xDefEq} to have vanishing mean $\expec{\x}=0$, and write its covariance matrix as 
\beq{Teq}
\T= \expec{\x\x^t}
=\left(
\begin{tabular}{cc}
$\C_0$&$\B$\\
$\B^t$&$\C_1$
\end{tabular}
\right).
\eeq
Modeling the probability distribution of $\x$ as a multivariate Gaussian, the mutual information between $\x_1$ and $\x_2$ is
\beq{Ieq}
I(\x_1,\x_2)={1\over 2}\log{|\C_1|\> |\C_2|\over |\T|},
\eeq
where we take $\log$ to denote the logarithm in base 2 so that information is measured in bits.

As mentioned, PCA elegantly decomposes the information content in a single random vector $\x$ into a mutually exclusive and collectively exhaustive set of information chunks corresponding to statistically independent numbers (eigenmodes coefficients) whose individual entropies add up to the total entropy. CCA generalizes this idea to mutual information, 
decomposing the total mutual information between $\x_1$ and $\x_2$ as a sum of the mutual information between a series of statistically independent pairs of numbers that are linear combinations of the two vectors, as summarized in Table~1 and in the following subsection.

\subsection{CCA implementation}

To do this, CCA first diagonalizes $\C_0$ and $\C_1$ as
\beq{EigenEq}
\C_1=\U_1\Lamb_1\U_1^t, \quad \C_1=\U_2\Lamb_2\U_2^t,
\eeq
where $\U_1$ and $\U_2$ are orthogonal matrices and $\Lamb_1$ and $\Lamb_2$ are diagonal, with the eigenvalues (which are non-negative up to numerical rounding errors) sorted in decreasing order.
It then constructs a prewhitening matrix
\beq{WhiteningEq}
\P\equiv
\left(
\begin{tabular}{cc}
$s(\Lamb_1)\U_1^t$&$\bzero$\\
$\bzero$&$s(\Lamb_2)\U_2^t$
\end{tabular}
\right),
\eeq
where the function $s(\lambda)\equiv\lambda^{-1/2}$, 
and a function of a diagonal matrix is defined by applying it to each diagonal element.
In many practical applications, some covariance matrix eigenvalues are near zero (and occasionally get evaluated as slightly negative due to numerical rounding errors), so below we implement CCA more robustly by instead defining 
\beq{fDefEq}
s(\lambda)=\left\{
\begin{tabular}{cl}
$0$				&$\quad$if $|\lambda|<\epsilon$\\
$\lambda^{-1/2}$	&$\quad$otherwise
\end{tabular}
\right.
\eeq
for some desired numerical precision floor $\epsilon$.
The matrix $\P$ transforms $\T$ into a block form
\beq{Teq2}
\T'
\equiv \P\T\P^t=
\left(
\begin{tabular}{cccc}
$\I$		&$\bzero$		&$\Q$		&$\bzero$\\
$\bzero$	&$\bzero$		&$\bzero$		&$\bzero$\\
$\Q^t$		&$\bzero$		&$\I$		&$\bzero$\\
$\bzero$	&$\bzero$		&$\bzero$		&$\bzero$\\
\end{tabular}
\right),
\eeq
where the zero-rows correspond to the eigenvalues that are so tiny that we round them to zero,
and the matrix $\Q$ can be interpreted as the Pearson correlation coefficients between the elements of the 
prewhitened vectors $\x_1$ and $\x_2$.
CCA now performs a singular-value decomposition (SVD) of $\Q$ \cite{eckart1936SVD}:
\beq{SVDeq}
\Q=\U\R\V^t,
\eeq
where the matrices $\U$ and $\V$ are orthogonal, $\R=\diag\{r_i\}$ is a diagonal matrix and 
$-1\le r_i\le 1$.
Defining 
\beq{Deq}
\D=
\left(
\begin{tabular}{cccc}
$\U^t$	&$\bzero$		&$\bzero$		&$\bzero$\\
$\bzero$	&$\bzero$		&$\bzero$		&$\bzero$\\
$\bzero$	&$\bzero$		&$\V^t$		&$\bzero$\\
$\bzero$	&$\bzero$		&$\bzero$		&$\bzero$\\
\end{tabular}
\right)\P,
\eeq
we can now transform our original covariance matrix $\T$ into the simple form 
\beq{Teq3}
\D\T\D^t=
\left(
\begin{tabular}{cccc}
$\I$		&$\bzero$		&$\R$		&$\bzero$\\
$\bzero$	&$\bzero$		&$\bzero$		&$\bzero$\\
$\R$		&$\bzero$		&$\I$			&$\bzero$\\
$\bzero$	&$\bzero$		&$\bzero$		&$\bzero$\\
\end{tabular}
\right).
\eeq
This means that when the random vector $\x$ is transformed into  
\beq{xprimeEq}
\x'=\left({\x'_1\atop\x'_2}\right)\equiv \D\x,
\eeq
there are no correlations between any elements of $\x'$ except between what we will term {\it ``principal pairs"} (to emphasize the analogy with principal components), matching elements in $\x_1'$ and $\x_2'$, 
which have a Pearson correlation coefficient 
$\expec{(\x_1')_i  (\x_2')_i}=r_i \in[-1,1]$.

Mutual information is independent of invertible reparametrizations of $\x_1$ and $\x_2$, so 
the mutual information from \eq{Ieq} now simplifies to 
\beqa{Ieq2}
I(\x_1,\x_2)
&=& I(\x'_1,\x'_2)
={1\over 2}\log{|\I|\> |\I|\over |\D\T\D^t|}=-{1\over 2}\log |\I-\R^2|\nonumber\\
&=&-{1\over 2} \sum_i \log(1-r_i^2)
\eeqa
if there are no numerically negligible eigenvalues.
If there are numerically negligible eigenvalues, the practically useful mutual information is given by this same formula,
since the corresponding eigenmodes are numerically untrustworthy.
This mutual information shared between $\x_1$ and $\x_2$ can be intuitively interpreted as stemming from a common source, as explained in Appendix~\ref{DistInterpSec}.




\subsection{One versus two latent representations}

As mentioned, dimensionality reduction is a popular approach to predicting the future of a time series $\z_i$ ($i=0,1, 2, ...$) in physics and other fields: the state vectors $\z_i$ are mapped into some lower-dimensional 
vectors $f(\z_i)$ by an invertible mapping $f$ that hopefully captures the most relevant information, after which the prediction is carried out in this latent space. This mapping can be either linear, such as in PCA or Independent Component Analysis \cite{hyvarinen2000independent}, or
non-linear as in autoencoders \cite{vincent2008extracting}, Generative Adversarial Networks \cite{goodfellow2014generative} or Contrastive Predictive Coding\cite{oord2018representation}.

For the special case of multivariate Gaussian probability distributions, the the formulas above imply that CCA provides the optimal dimensionality reduction, with the twist that the mappings into the latent space should generally be different for the predictor vector and the predicted vector:
we can define the CCA dimensionality reduction as the mapping 
\beq{CCAdefEq1}
\u\equiv\F\x,\quad \vv\equiv\G\y
\eeq
into the latent space $\Re^k$, where 
\beq{CCAdefEq2}
 \F\equiv \PP_k \U^t\Lamb_1^{-1/2}\U_1^t, \quad \PP_k \F\equiv\V^t\Lamb_2^{-1/2}\U_2^t
 \eeq
and the projection matrix $\PP_k$ simply picks the first $k$ elements of vector following it.
The CCA construction above is readily seen to imply that 
 $$I(\u, \vv) \ge I[f(\x),g(\y)]$$ for any functions 
 $f$ and $g$ mapping $\x$ and $\y$ into $\Re^k$, 
 and $I(\u, \vv)=I(\x,\y)$ when the dimensionality is not reduced below that of both $\x$ and $\y$.
 
 Note that generically, $\F\ne\G$, even for time series where $\x$ and $\y$ live in the same space. The following theorem shows that this is a feature, not a bug

{\bf Theorem:} A single latent representation sometimes underperforms two separate ones, capturing less mutual information in some given number of variables.

\begin{figure}[phbt]
\centerline{\includegraphics[width=88mm]{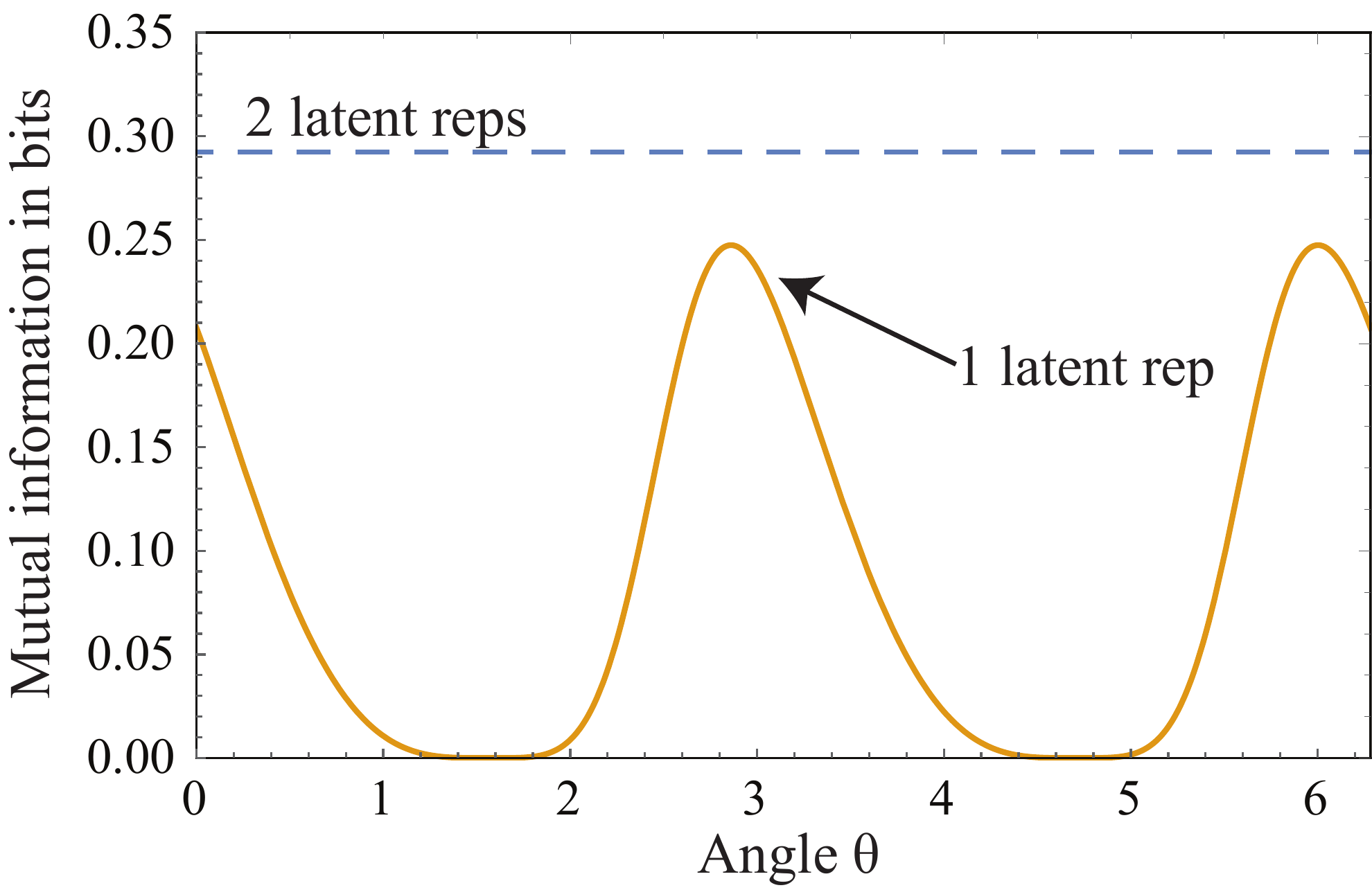}}
\caption{Counterexample showing a single latent representation underperforming two separate ones.
\label{CounterexampleFig}
}
\end{figure}

{\bf Proof: }
A  simple counterexample (to the hypothesis that a single representation is equally good) is  that 
is provided by four random variables $x_1$, $x_2$, $y_1$, $y_2$ with unit variance and no correlations except that $\expec{x_1 x_2}=\expec{x_1 y_1}=1/2$.
The CCA described above shows that the mutual information can be entirely captured by a single principal pair $\{u_1,v_1\}\equiv\{2x_1-x_2,y_1\}$; specifically, 
\beq{CounterExampleIeq}
I(\x,\y)= I(2x_1-x_2,y_1)={\log{3\over 2}\over\log 4}\approx 0.29\>\hbox{bits}.
\eeq
If we instead transform both $\x$ and $\y$ using a single latent representation, then
the maximal mutual information we can attain from a single principal pair is smaller:
\beqa{CounterExampleEq}
I(\x,\y)&>&\max_\w I(\w\cdot\x,\w\cdot\y)\nonumber\\
&=& \max_\theta I(x_1\cos\theta+\x_2\sin\theta,y_1\cos\theta+\y_2\sin\theta)\nonumber\\
&\approx&0.247\>\hbox{bits},
\eeqa
as can be seen in \fig{CounterexampleFig}. Here we have without loss of generality taken $\w$ to be a unit vector $\w=\{\cos\theta,\sin\theta\}$, since the mutual information above is invariant under rescaling  $\w$.

As mentioned, most published work uses merely a single latent representation for time series prediction.
This is clearly not optimal for the general case, since we have just proven that it is not even optimal for the simple case of multivariate Gaussian distributions. 
But does this suboptimality really matter in practice? 
The suboptimality is seen to be only 0.05 bits in \fig{CounterexampleFig}), so it is natural to ask whether 
a single latent representation $f$ is generically close to optimal, or whether the further improved from adding a second reporesentation $g\ne f$ provides large enough an improvement to be worth the extra complication. We address this questions in the next section by a practical application. 

\section{Example: Coupled harmonic oscillators with dissipation and thermal noise}
\label{OscilatorSec}

\begin{figure}[phbt]
\centerline{\includegraphics[width=88mm]{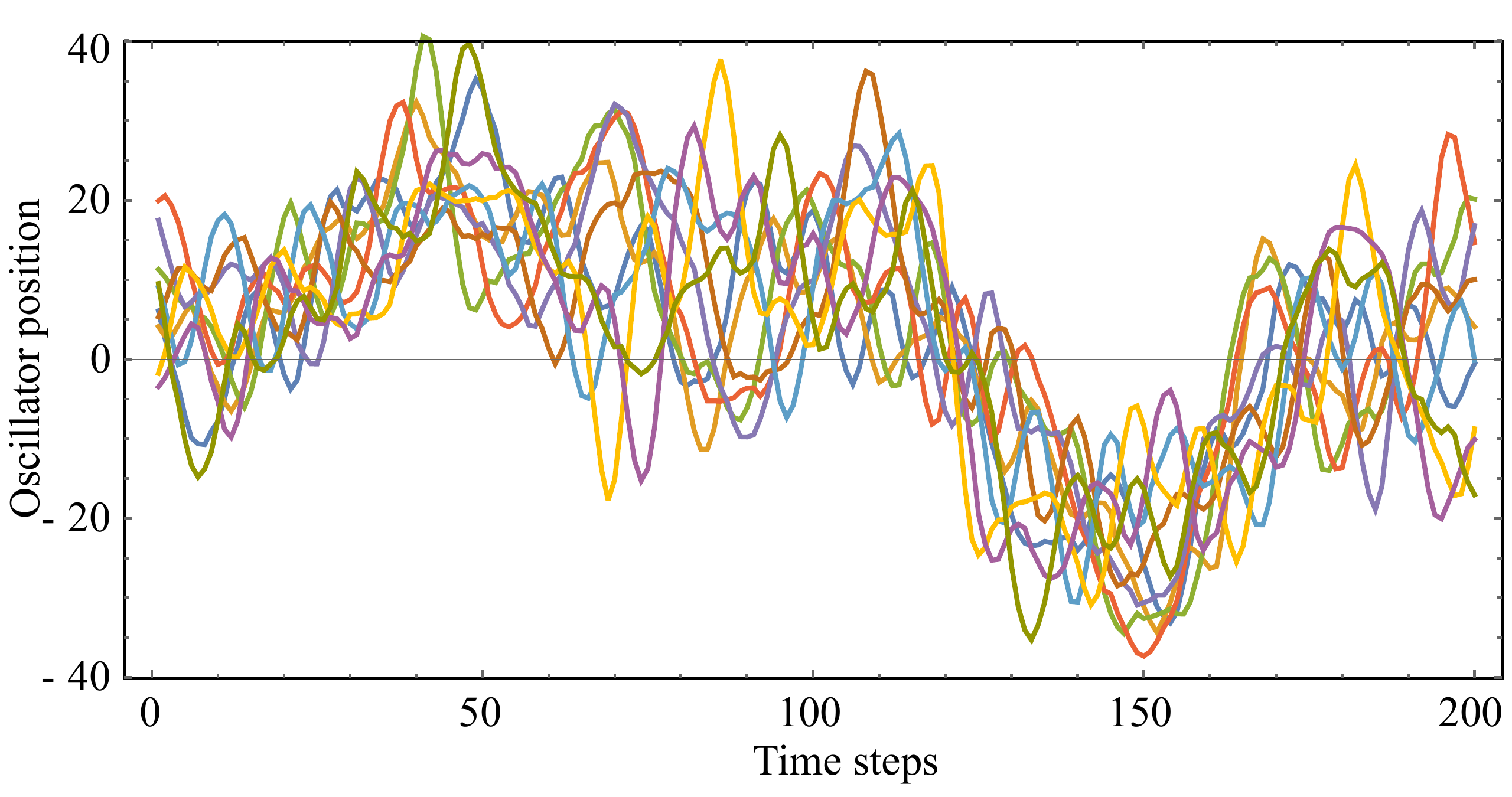}}
\caption{Sample evolution of our ten coupled harmonic oscillators, damped by friction and perturbed by thermal noise. \label{evolutionFig}
}
\end{figure}

To better compare the predictive abilities of CCA with other approaches, let us now consider 
the physics problem of predicting the future state of a set of $n$ coupled 1-dimensional harmonic oscillators that are damped by friction and perturbed by random thermal noise.
We group the positions and momenta of the oscillators into the $n$-dimensional vectors $\q$ and $\p$, 
which we in turn group into a single $2n$-dimensional state vector $\z$.
We set all masses equal to unity, so $\dot{\q}=\p$, 
and take the laws of motion to be $\dot\p=-\K\q-\GG\p$ for some positive semidefinite spring matrix $\K$ and friction matrix $\GG$.
This means that we can write 
\beq{DynamicsEq}
\dot\z=\left({\dot\q\atop\dot\p}\right)=
\B\left({\q\atop\p}\right)=\B\z,
\quad
\B\equiv 
\left(
\begin{tabular}{cc}
$\>\>\>\bzero$&$\I$\\
$-\K$&$\GG$
\end{tabular}
\right),	
\eeq
which has the solution 
$$\z(t)=e^{\B t}\z(0).$$
All eigenvalues of $\B$ are negative, so to prevent $\z(t)$ from simply decaying toward zero, we add random Gaussian noise of standard deviation $\sigma$ to each position at every time step $\tau$, 
Defining $\z_i\equiv\z(\tau i)$, we can thus rewrite our time-evolution as a Markovian autoregressive process:
\beq{AutoregressiveEq}
\z_{i+1}=\A\z_{i}+\n_i,
\eeq
where 
$\A\equiv e^{\B\tau}$, $\expec{\n_i}=\bzero$, $\expec{\n_i\n_j^t}=\delta_{ij}\Sig$,
$\Sig_{kl}=\delta_{kl}\sigma^2$ if $k\le n$ and zero otherwise (the standard deviation is $\sigma$ for position noise, zero for momentum noise). 
This random process will eventually converge to a stationary state whose probability distribution is time-independent, since all eigenvalues of $\A$ have magnitude below unity, so that memory of the past gets exponentially damped over time.
\Fig{evolutionFig} shows an example for $n=10$ oscillators arranged in a circle. Here and below we use time step $\tau=1$, noise level $\sigma=1$, friction matrix $\GG=\gamma\I$ with $\gamma=0.05$, and
spring matrix $\K$ corresponding to nearest-neighbor coupling $\alpha^2=0.2$ and self-coupling $\omega^2=0.01$, \ie, $K_{ij}=\omega^2/2+\alpha^2$ if $i=j$, $K_{ij}=-\alpha^2/2$ for nearest neighbors, and $K_{ij}=0$ otherwise.

Once stationarity has been attained, the mean $\expec{\z}$ vanishes and \eq{AutoregressiveEq} implies that the time-independent covariance matrix
 $\C\equiv\expec{\z_i\z_i^t}$ satisfies $\C=\A\C\A^t+\Sig$. 
This is known as the Lyapunov equation, and is readily solved for $\C$ by special-purpose techniques or, rapidly enough, by simply iterating it to convergence.

We are interested in using the state vector $\z_i$ to predict the subsequent state $\z_{i+k}$.
Arranging these two $2n$-dimensional vectors into a single $4n$-dimensional vector $\x$, 
we can now compute the 2-time covariance matrix of \eq{Teq} by 
iterating \eq{AutoregressiveEq} $k$ times:
\beq{Teq2}
\T^{(k)}= \expec{\x_i\x_{i+k}^t}
=\left(
\begin{tabular}{cc}
$\C$&$(\A^k\C)^t$\\
$\A^k\C$&$\C$
\end{tabular}
\right),
\eeq

\begin{figure}[phbt]
\centerline{\includegraphics[width=88mm]{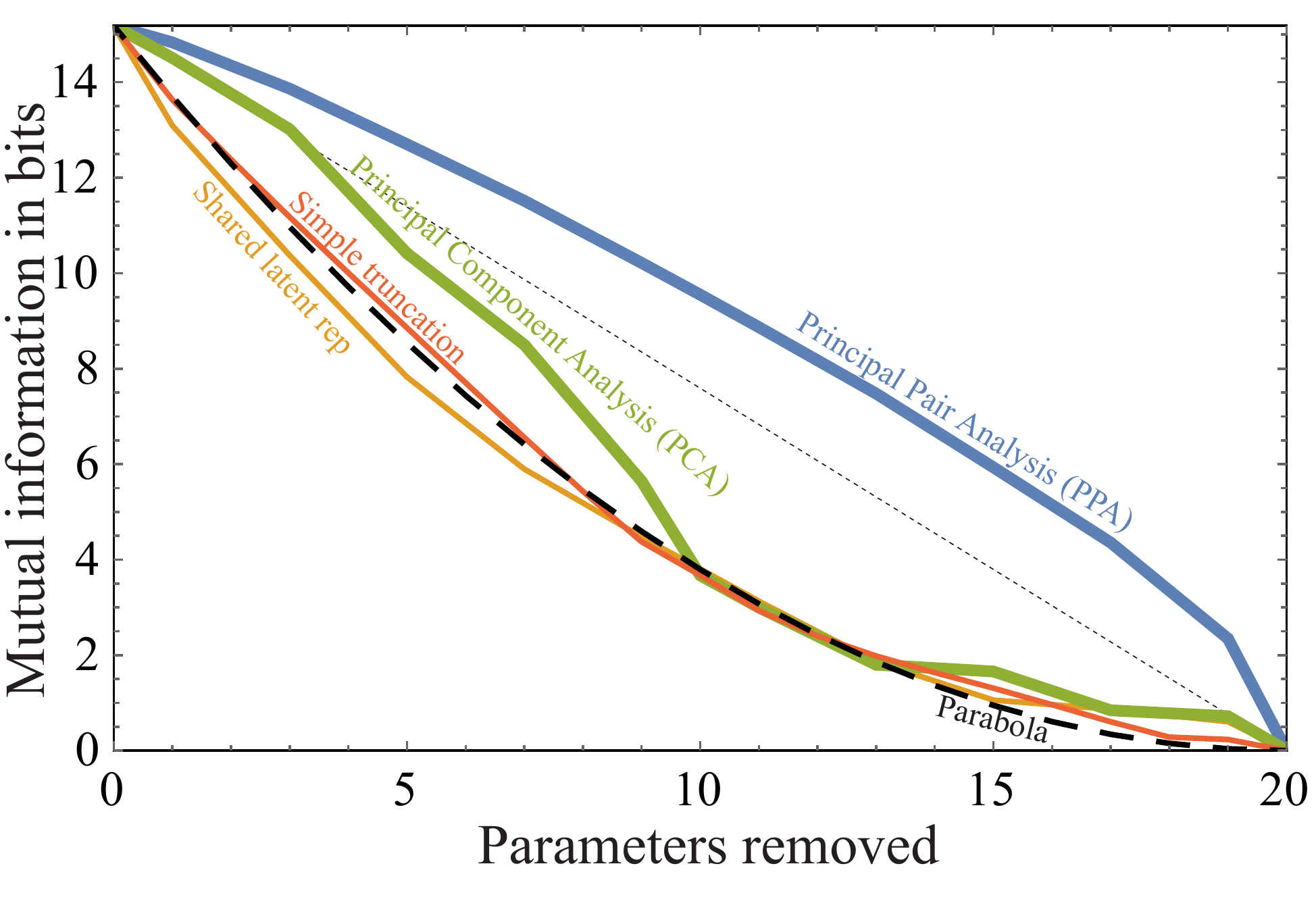}}
\caption{Pareto frontier for different dimensionality reduction methods
$\x\mapsto\u$, $\y\mapsto\vv$, showing the mutual information $I(\u,\vv)$ as a function of the dimensionality reduction.
\label{informationFig}
}
\end{figure}

\Fig{informationFig} shows the mutual information $I(\x_i,\x_{i+10})$
between the the state of our harmonic oscillators and their state $10$ time-steps later, as a function of the amount of dimensionality reduction performed. 
The CCA curve plotted is by definition the Pareto frontier for the tradeoff between compression and information retention, \ie, the the maximum amount of information that can be collectively retained in a given number of pairs.  The CCA curve is seen to be concave because all Pareto frontiers by construction have non-positive second derivative.

Three other dimensionality reduction methods are also plotted for comparison, and are all seen to perform significantly worse than CCA. These all use the same latent representation for both 
$\x_i$ and $\x_{i+10}$. The PCA curve keeps the top principal components, while the ``simple truncation" curve simply retains the first elements of the vectors $\x_i$ and $\x_{i+10}$.
The ``shared latent rep" curve uses the CCA-matrix $\F$ to compress $\x$, and uses the same matrix $\F$ (rather than $\G$) to compress $\x_{i+10}$.

If each pair of numbers $\{u_i,v_i\}$ contained the same fraction of the total mutual information, 
then the Pareto frontier would be a straight line. The performance of the non-CCA methods is seen to be even worse in our example, closer to a parabola (dashed line). This is what one expects from any method where each number $u_i$ contains a random fraction $1/N$ of the total information, and the same holds for each number $v_i$:
then the mutual information $I(u_i,v_j)=1/N^2$, and
the information fraction shared by $N'$ numbers $u_i$ and $N'$ numbers $v_i$
is $(N'/N)^2$ – a parabola.




\section{Conclusion}
\label{ConclusionsSec}

It is often useful to map data vectors into a lower-dimensional latent space, retaining only the information of interest, as summarized in \fig{ComparisonTable}.
For linear mappings $f$ and $g$, the natural generalization of PCA is CCA,  which distills two random vectors $\x$ and $\y$ 
into linear transformations $\u=\F\x$ and $\vv=\G\y$ such that all components of both vectors are uncorrelated, except that matching ``principal pairs" have correlation $\r_i \in[-1,1]$.
For Gaussian random vectors, CCA conveniently decomposes the total mutual information between $\x$ and $\y$ as the sum of the mutual information $\log(1-r_i^2)$ between these principal pairs.
Retention of only the $k$ most informative pairs thus 
falls on the Pareto frontier of optimal dimensionality reduction.

There is strong current interest in how to best generalize this to nonlinear mappings optimized for non-Gaussian random vectors. Most recent work for non-linear time-series prediction (see  \cite{oord2018representation} and references therein) focuses on the special case $f=g$.  
As we have explored in detail, this can be far from optimal, even in the linear case. 
In the linear case, the reason why two different latent representations $\F$ and $\G$ are better than one ultimately traces back to the fact that the SVD in \eq{SVDeq} produces different matrices  $\U\ne\V$ when $\Q$ is asymmetric. For our harmonic oscillator example (or any physics time series whatsoever, for that matter), this asymmetry 
corresponds to time reversal asymmetry: the operation to predict the future state from the past is not the same as that predicting the past from the future. 
In our example, this asymmetry can be eliminated by ignoring all momentum information: 
Repeating CCA to predict $\q_{i+k}$ from $\q_i$ (as opposed to working with $\x_i$ which includes both $\q_i$ and $\p_i$), we obtain $\F=\G$.

\begin{figure}[phbt]
\centerline{\includegraphics[width=88mm]{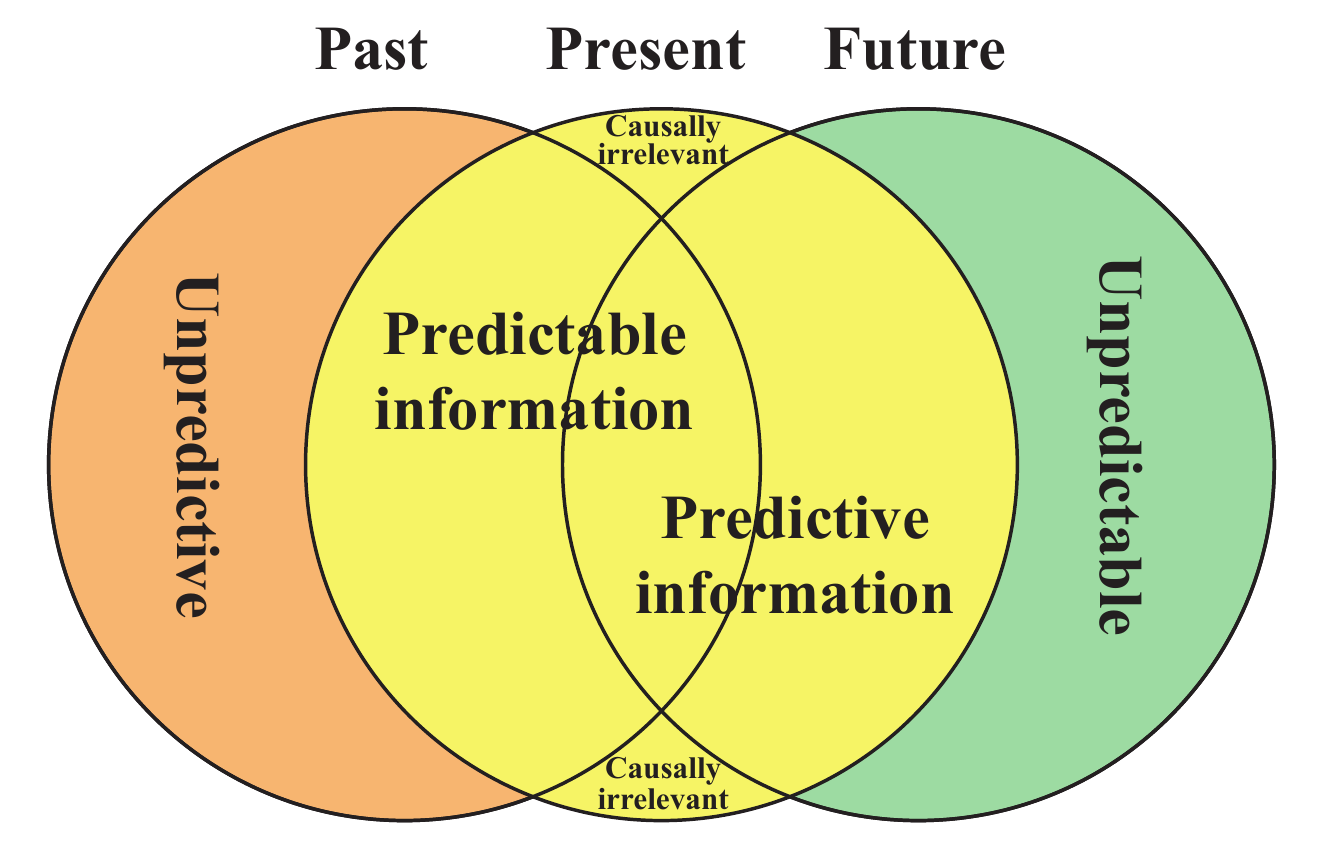}}
\caption{The information in a system (represented by a circle) that helps predict its future state may differ from the information that is predictable from its past state; the remainder is causally irrelevant, containing no information about either past or future.
\label{VennFig}
}
\end{figure}

A natural conjecture for future work to investigate is that this generalizes to non-Gaussian time series where the optimal dimensionality reduction is nonlinear: that a single latent representation suffices for reversible Markov chains and other reversible processes, while a pair of representations performs better for processes that are truly different in reverse, for example text, speech,  music or out-of-equilibrium physical systems.

\Fig{VennFig} motivates this conjecture: when the evolution of a dynamical system is not time-reversible, 
then the information about its state that is useful for predicting what will happen is often different from that which is useful for predicting what happened. 
Consider, for example, a table with an orange and a pencil balanced on its tip, both isolated from the rest of the world and seemingly at rest. If we are interested in predicting the future, we should pay more attention to the orange, since it will stay put while the pencil will tip over in less than 30 seconds in a direction that we cannot predict, as quantum-mechanical fluctuations get amplified by gravity. 
If we are interested in inferring the past, we should instead focus on the pencil, which was almost certainly at least as balanced then as it is now. The orange, on the other hand, is in a stable attractor state: it could either have just sat there, or slid/rolled and come to rest due to dissipation. 
In physics examples such as these, whether information is predictable, predictive or both thus depends on attractor dynamics and  Lyapunov exponents: degrees of freedom in stable equilibria (with negative Lyapunov exponents) are predictive but unpredictable, while those in unstable equilibria (with positive Lyapunov exponents) are unpredictive but predictable. 

The causally irrelevant information, that helps with neither, should obviously be discarded in latent representations; if most of the information is in this category, then 
non-causal approaches such as PCA, ICA and nonlinear autoencoders may mistakenly retain much of this information if it is easy to distill into a small number of variables.
Renormalization in physics provides an example of a single latent representation where the vast majority of the information (typically information on very small scales/high frequencies) is discarded, while both the longer-term predictive and predictable information is retained. 
In other cases, the predictable and predictive degrees of freedom parts of the information may be closer to disjoint, in which case switching to two separate representations offers the opportunity of cutting the latent space dimensionality almost in half. 
An interesting challenge for future work is therefore  to explore whether approaches such as 
\cite{oord2018representation} can be further improved by using more than one latent representation, or by developing a single one that better optimizes for both predictiveness and predictability.


\bigskip

{\bf Acknowledgements:} The author wishes to thank Tailin Wu for helpful comments, and for suggesting \fig{VennFig} and the idea behind it.
This work was supported by The Casey and Family Foundation, the Foundational Questions Institute, and by Theiss Research through TWCF grant \#0322. 



\appendix

\section{Distillation interpretation} 
\label{DistInterpSec}

The following theorem allows an intuitive interpretation of correlation as stemming from a common source. 
\smallskip

{\bf Theorem:}
Any random vector pair can be decomposed as the sum of a perfectly correlated part (``signal") and a perfectly uncorrelated part (``noise"):
$$\left({\x_1\atop \x_2}\right)=
\left({\A_1\atop \A_2}\right)\s + \left({\n_1\atop \n_2}\right),
$$
where 
$\expec{\s\s^t}=\I$, $\expec{\n_1\n_2^t}=\bzero$, $\expec{\s\n_1^t}=\bzero$ and  $\expec{\s\n_2^t}=\bzero$.

{\bf Proof:}
Writing the last equation as $\x=\A\s+\n$,
we have $\expec{\n}=\bzero$,  and $\expec{\s\n^t}=\bzero$. Define $\Sig\equiv\expec{\n\n^t}$.
If $\x_1$ and $\x_2$ have the same length and no eigenvalues of $\C_1$ or $\C_2$ vanish, then define 
\beq{AdefEq}
\A\equiv \P^{-1}\D^{-1}\left(
\begin{tabular}{cc}
$\R^{1/2}$\\
$\R^{1/2}$
\end{tabular}
\right),
\eeq
\beq{SigmaDefEq}
\Sig\equiv\D^{-1}\left(
\begin{tabular}{cc}
$\I-\R$&$\bzero$\\
$\bzero$&$\I-\R$
\end{tabular}
\right)\D^{-t}.
\eeq
It follows that 
\beqa{ProofEq}
\expec{\x\x^t}&=&\A\expec{\z\z^t}\A^t+\expec{\n\n^t}+\A\expec{\z\n^t}+\expec{\n\z^t}\A^t\nonumber\\
&=&\A\A^t+\Sig=
\P^{-1}\D^{-1}
\left(
\begin{tabular}{cc}
$\R$&$\R$\\
$\R$&$\R$
\end{tabular}
\right)\D^{-t}\P^{-t}+\Sig\nonumber\\
&=&
\P^{-1}\D^{-1}\left(
\begin{tabular}{cc}
$\I$&$\R$\\
$\R$&$\I$
\end{tabular}
\right)\D^{-t}\P^{-t}\nonumber\\
&=&\P^{-t}\D^{-1}(\D\T\D^t)\D^{-t}\P^{-t}\nonumber\\
&=&\P^{-t}\T'\P^{-t}=\T
\eeqa
as required, where we used \eq{Teq3} in the penultimate step.
The case with vanishing eigenvalues and/or unequal length of $\x_1$ and $\x_1$ follows straightforwardly from zero-padding appropriately. 

We can thus interpret $\s$ as the distillation of all the correlated information in $\x_1$ and $\x_2$, which is shared but diluted by noise.

\bibliography{latent}

\end{document}